\documentclass{ieeeaccess}
\usepackage{tabularx,booktabs, longtable}
\usepackage{supertabular}
\usepackage{booktabs}
\usepackage{cite}
\usepackage{amsmath,amssymb,amsfonts}
\usepackage{algorithmic}
\usepackage{graphicx}
\usepackage{textcomp}
\def\BibTeX{{\rm B\kern-.05em{\sc i\kern-.025em b}\kern-.08em
    T\kern-.1667em\lower.7ex\hbox{E}\kern-.125emX}}
\newcolumntype{C}{>{\centering\arraybackslash}X} 
\begin{document}
\history{}
\doi{}

\title{Students Success Modeling: Most Important Factors}

\author{\uppercase{Sahar Voghoe}\authorrefmark{1,2}, 
\uppercase{James M. Byars\authorrefmark{2}, Scott Jackson King\authorrefmark{2},Soheil Shapouri\authorrefmark{3},Hamed Yaghoobian\authorrefmark{4},Khaled M. Rasheed\authorrefmark{3}, Hamid R. Arabnia\authorrefmark{3}}}
\address[1]{University of Georgia, Athens, GA, USA (e-mail: voghoei@uga.edu)}
\address[2]{Carl Vinson Institute of Government}
\address[3]{University of Georgia, Athens, GA, USA}
\address[4]{{Muhlenberg College, Allentown, PA, USA}}

\tfootnote{This work was supported in part by Carl Vinson Institute of Government}

\markboth
{Author \headeretal: Preparation of Papers for IEEE TRANSACTIONS and JOURNALS}
{Author \headeretal: Preparation of Papers for IEEE TRANSACTIONS and JOURNALS}


\begin{abstract}
The importance of retention rate for higher education institutions has encouraged data analysts to present various methods to predict at-risk students. The present study, motivated by the same encouragement, proposes a deep learning model trained with 121 features of diverse categories extracted or engineered out of the records of 60,822 postsecondary students. The model undertakes to identify students likely to graduate, the ones likely to transfer to a different school, and the ones likely to dropout and leave their higher education unfinished. This study undertakes to adjust its predictive methods for different stages of curricular progress of students. The temporal aspects introduced for this purpose are accounted for by incorporating layers of LSTM in the model. Our experiments demonstrate that distinguishing between to-be-graduate and at-risk students is reasonably achievable in the earliest stages, and then it rapidly improves, but the resolution within the latter category (dropout vs. transfer) depends on data accumulated over time. However, the model remarkably foresees the fate of students who stay within the school for three years. The model is also assigned to present the weightiest features in the procedure of prediction, both on institutional and student levels. A large, diverse sample size along with the investigation of more than one hundred extracted or engineered features in our study provide new insights into variables that affect students’ success, predict dropouts with reasonable accuracy, and shed light on the less investigated issue of transfer between colleges.More importantly, by providing individual-level predictions (as opposed to school-level predictions) and addressing the outcomes of transfers, this study improves the use of ML in the prediction of educational outcomes.
\end{abstract}


\titlepgskip=-15pt

\maketitle

The growing importance of the role of postsecondary educational institutions in modern societies has given significance to the event of graduation, a recognition of the ultimate educational achievement of a postsecondary student. The importance of graduation lies in the fact that it is the most formal and straightforward criterion that functions as a necessary condition to let people officially enjoy the benefits of their institutional education. 

As much as graduation symbolizes educational success in general, dropping out of a program, on the contrary, signifies the failure of a student as well as that of the concerned educational institution. Dropping out possibly denotes unwise management of time, money, and energy, which can potentially hurt a student's confidence, distance them from their intended goals, and adversely affect their professional positions and income in the future. At the same time, it hurts the retention rate of the institution at which the student was admitted. This, consequently, threatens the reputation of the institution, which may limit its access to financial resources, including tuition fees, donations, and state or federal support. Also, at the national level, the increased attrition rate represents an insecure picture of educational investment both for students and the involved stakeholders with economic motivations in addition to hurting the efficiency of national resource management. 

In addition to termination of education, which we refer to as ``dropout'' in this article, transferring is one of the common scenarios in which the attendance of a student at a school may end up. Although transfer students endanger the retention rate of the schools they transfer from, a great number of them land in another school and complete a program, which serves the retention rate of their destinations. In other words, schools that are not able to dissuade students from transferring out lose a part of the effort they have made and the money they have spent on the schools that attract transfer students. 

Machine learning techniques have been proven to be able to prevent dropping out, to an effective extent, by identifying at-risk students for advisory and supplementary treatments. Studies have also shown that machine learning models can single out the features that contribute the most to this undesired outcome, which is a concern of many policymakers. However, previous studies suffer from several methodological issues: 

\begin{enumerate}
    \item They have not taken transferring into consideration as an independent contributor to attrition, while, since the motivations for transferring are not necessarily the same as those of dropping out, they may require different treatments to be prevented.
    
    \item Previous studies typically are not able to establish a cause-effect relationship. Therefore, if the feature they single out as the main contributor is a co-effect of the phenomenon under investigation, it will not offer much information regarding the factors that may have caused the problem and, accordingly, need to be treated.
    
    \item Although the models presented in previous studies are able to identify the features that generally weigh in the process of predicting at-risk students, they have not been assigned to deliver this task for each student at the individual level. Providing the latter information may decisively assist advisors and instructors in personalizing remedial or motivational packages to achieve a better effect and efficiency.

    \item Previous studies on this subject either confine their attention to dropout cases that occur within the early semesters of enrollment or consider all cases regardless of the stages of curricular accomplishments. The former approach ignores later dropouts, although the investment accumulated over several semesters lost due to a late dropout is much more than what an early dropout dissipates. The latter approach blurs the resolution advisors' need to specify the components of the packages intended to improve at-risk students during different semesters.  

\end{enumerate}

The present study, motivated to address the aforementioned shortcomings, proposes a deep learning model trained with 121 features of diverse categories extracted or engineered out of the records of 60,822 postsecondary students from 2006 to 2019. While this large sample size makes the model more realistic, the diversity and abundance of features make it more likely to identify influential causal factors that educators can further manipulate. This model aims to identify students likely to graduate from a university of interest, the ones likely to transfer to a different school and graduate, the ones who transfer but never graduate, and the ones likely to drop out and leave their higher education unfinished. The present study also accepts the under-investigated challenge of predicting transfer students. While most studies related to the issue of transfer (e.g.,\cite{freitas2020iot}) have ignored different outcomes of transfers, we have investigated the predictors of transferring down (to a lower-ranked college) and transferring up (to a better school.) The predictions of this model will be made across different stages of education. The model is also assigned to present the most impactful features with respect to each of the three categories, both on institutional and individual levels.

\section{Background}
Among the studies that inspired us, two classes should be mentioned here: 1) the ones whose objective was to predict students who would drop out (either from a course or form a program) and 2) the ones that undertake to predict the final performance of a student in a course. 

As a representative of the first class, \cite{-56} should be cited. This study is also a remarkable example of studies that deal with online courses. It extracted data from the MOOC (Massive Open Online Courses). The authors explored the performance of a deep fully-connected feed-forward neural network to detect students likely to drop out. The model achieved an accuracy of 90.20\%. Another example of this class is the study of \cite{-40}, which was one of the most sophisticated projects in this class. One of the most interesting features of this project is that, as a part of its preprocessing, courses within similar academic fields were clustered in the following categories: system engineering, mathematics, physics, language, management, and biology. Finally, Decision Trees, Logistic Regression, and Naïve-Bayes were tried, among which the former had the best performance (with AUC of 94\%). The model recognized the average of grades earned in the group of “systems engineering” courses and the accumulative grade point average among the most effective features. \cite{berens2019early} also proposed an early detection system that predicts students' dropout. For this purpose, this study used the AdaBoost Algorithm to combine regression analysis, neural networks, and decision trees. The approach presented in this study was later replicated and effectively modified by \cite{wagner2020accuracy}. Finally, in this class, we cite \cite{sandoval2020early}, that illustrated the triumph of Artificial Neural Networks over Logistic Regression. In this study, the optimized neural network model, which had a 4-7-1 architecture, returned a classification accuracy of 0.768 and the area under the ROC curve of 0.795.

As inspiring examples for the second class, which aimed to predict the performance of students in a course or a group of courses, we should mention the following. \cite{-43} employed and compared Support Vector Machines (SVM), Random Forests, Rule-based Classifier, Trees (J48), PART, IB1, and Naïve-Bayes to predict the final grades of students right at the beginning of a semester. The data this study has collected include social networking metrics indexed according to the method of Pajek. \cite{-42} focused on predicting students’ failure in introductory programming courses. Decision Tree (J48), Support Vector Machine, Neural Network, and Naive Bayes, with and without fine-tuning were the models this study explored. The study claims that Decision Tree presented the highest effectiveness. \cite{-2018} covered a large number of students. The study asserts that the performance of Random Forest was the best among all methods the authors tried. It has also contributed to the prediction of at-risk students more than what it did to the identification of students who would earn high grades. In addition, it states that the most significant variable in the procedure of prediction, as already suggested by the literature \cite{-38}, was the Cumulative Grade Points Average. \cite{saheed2018student} committed to gathering data from administrative channels. This study adopted and compared Iterative Dichotomiser 3, C4.5, and Classification and Regression Tree (CART) to predict students' performance, and finally indicated the superiority of C4.5 and Simple CART in terms of accuracy (both with 98.3\% accuracy). To predict student’s performance in a course,\cite{-52} introduced a new algorithm called GritNet, which is based on Bi-directional Long Short-term Memory (BiLSTM). According to this study, GritNet does not require feature engineering. Ye Mao\cite{-57}, is significant for employing an automatic skill discovery method (SK) that was intended to reduce the expert's dependency of the model. The study claims, “BKT and BKT+SK outperformed the others on predicting post-test scores, whereas LSTM and LSTM+SK achieved the highest accuracy, F1-measure, and area under the ROC curve (AUC) on predicting learning gains \cite{-57}.”

Selecting the features with significant influence on educational performance and persistence, we were inspired by researchers who addressed the impact of a specific feature or a group of features in this respect. Below, we cite some of them.

\cite{britt2017student} found that financial stress and students' self-reported debt contribute to the likelihood of discontinuing education. \cite{kim2018sustainable} also claimed that the cost and burden of education for students as well as financial resources have consequential effects on dropout rates. The effects of another factor like gender on college persistence and completion have been also documented \cite{weeden2020pipeline}.

While \cite{westrick2015college,weston2019predicting} compared the significance of high school achievements and ongoing university achievements, \cite{stenton2021fine} addressed the effects of distance from home and Gini index value of home area on attrition. \cite{niessen2016predicting} also emphasized the contribution of academic scores to the prediction of educational success after the first year. In the same light, \cite{gershenfeld2016role} used a set of logistic regression models to establish low first-semester GPA as an indicator of failure to complete an undergraduate program in six years. Finally, \cite{davidson2017initial} showed the importance of initial academic momentum in the first year of college for academic persistence and earning a degree.

\section{Data Gathering}
The task of our proposed model is to predict the educational success of students in a postsecondary institution. The measures will be scaled in four categories, namely ``graduate'', ``transfer'', ``transfer to-be-graduate'', and ``dropped-out'', defined as follows:
\begin{itemize}
\item Graduate: a student who has been awarded a degree by the institution in the last semester of enrollment.

\item Transfer: a student who has not been awarded a degree by the last semester of enrollment in an institution, but has reappeared in the postsecondary education system within two years without any record of graduation. 

\item Transfer to-be-graduate: a student who has not been awarded a degree by the last semester of enrollment in an institution, but has reappeared in the postsecondary education system within two years, and then, graduated from the destination institute.

\item Dropped-out: a student who has not been awarded a degree by the last semester of enrollment in an institution, and has not reappeared in the postsecondary educational system within two years.
\end{itemize}

\begin{figure}[hbt!]
\centering
    \includegraphics[width=0.5\textwidth]{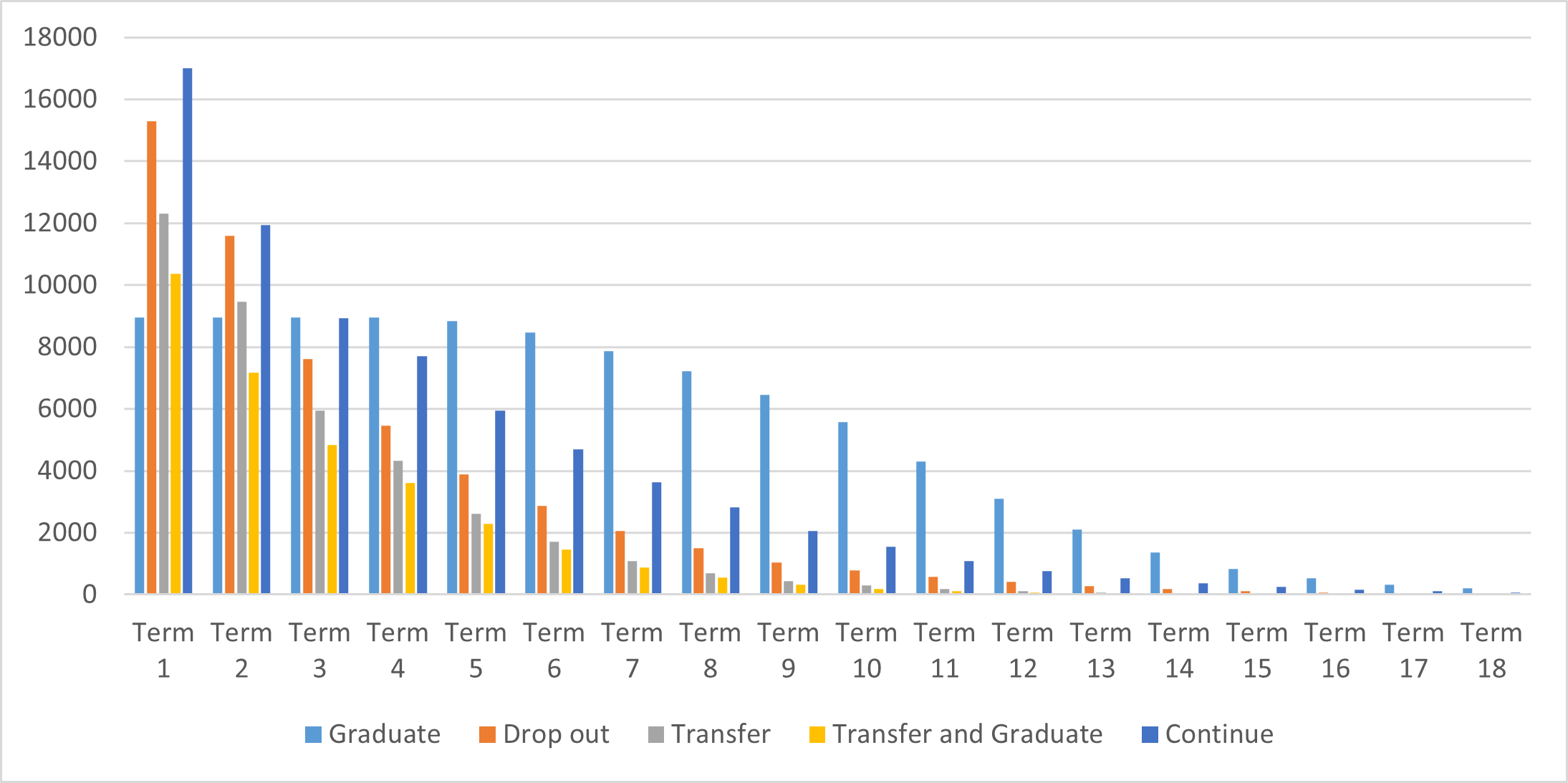}
    \caption{Labels}
    \label{fig:LabelsDist}
\end{figure}

Hereafter, we call each of these categories a ``label''. Naturally, students who are not labeled with any of these categories are supposed to be continuing their education at the same institution, and therefore, have no significance for training our model. However, whenever the model is to predict the immediate fate of students after each semester of enrollment, ``continuation'' should be introduced to the model as the fifth label. 

For the present phase of our project, we found enough reason to consider 121 features as the factors that may correlate with the distribution of the aforementioned labels. Below, we briefly introduce these features within the following classes:

\begin{itemize}
    \item Personal data: e.g., age, race, ethnicity, gender, sex, job, army employment, risk of homelessness, etc.
    \item Family data: e.g., family size, parents age, orphanage, parents' financial data, number of dependents, parents' education, family's academic background, number of studying or graduate siblings, family income and loans, etc.
    \item Pre-college data: e.g., high school subject grades, high school accumulative GPA, the scores of SAT, COM, and CAT, etc.
    \item Financial data: e.g., loans like PELL and HOPE (applied or approved), scholarships and financial awards, tuition fees, parking, and meal expenses, etc. 
    \item Academic data: credits transferred, credits attended, credits passed, GPA (accumulative and the last semester), class size, withdrawals, etc.
    \item Online courses: number of credits, grades, forum, assignment submissions, quiz grades, etc. 
    \item Online activity index: duration of online activities, the number of discussions attended, etc. 
\end{itemize}

Table \ref{tab:Compare} presents a full list of these features. These features along with the aforementioned labels define the dimensions of the domain with which our model trains. In order to maintain the values representing these features and labels, we extracted data from three resources that will be elaborated in the following subsection. 

\begin{table*}[htbp]
    \centering
    \includegraphics[width=2\columnwidth]{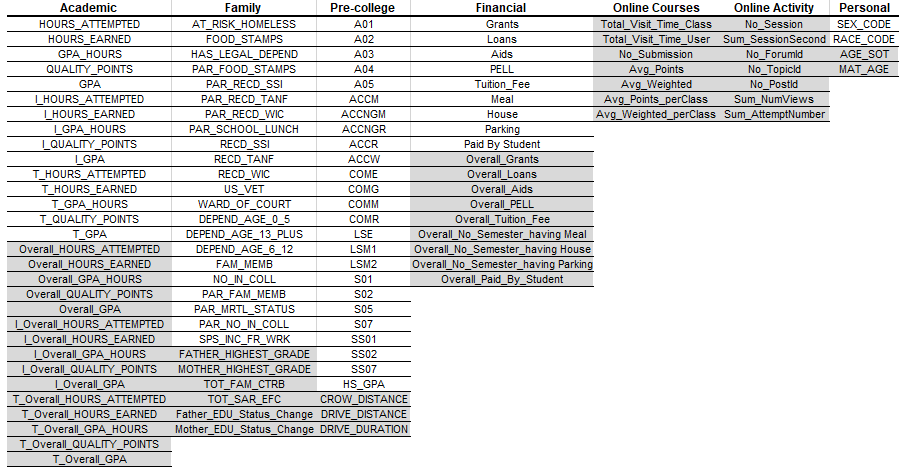}
    \caption{The Features in each family and the gray once are engineered features.}
    \label{tab:Compare}
\end{table*}

\subsection{Resources}
The variety of features that we mentioned earlier required us to gather data from three different resources, namely Banner, Brightspace, and the National Student Clearinghouse. As explained below, none of these resources would solely cover all categories of the intended features. For further operations, we stored the data gathered from these resources in a central database. 

\subsubsection{Banner} 
Banner is an Oracle-based Enterprise Resource Planning (ERP) software that assists colleges and universities in recording and maintaining data. Most of the University System of Georgia (USG) members use this software for several facilities. Among these facilities, we used its data storage, which allowed us to gather various types of information, including ones under the following classes: personal data (4 features), family data (28 features), pre-college data (28 features), financial data (18 features), and academic data (30 features). This data can be as specific as the information regarding class assignments and activities for all academic courses.

One of the most interesting groups of data stored by Banner is the information that FAFSA (Free Application for Federal Student Aid) forms provide. For some features related to family and financial status, these forms constitute the only source of data. However, the fact that FAFSA forms are provided only when students apply for federal financial aids leaves us with many missing values for these features. Out of all the students whose data we have collected for our study, 42911 applied for funds through FAFSA at least once, while 17911 students never did so. 

Banner also provides us with the graduation status of students and their last date of enrollment, which we used to specify the labels we mentioned earlier. Banner's data also aids to identify the two following categories of students that should be excluded from our study: 

\begin{enumerate}
    \item Military students. We exclude their data because of two sorts of anomalies their records would introduce to our study: firstly, many of their records are sealed; secondly, special rules of attendance they follow allow them to stay unenrolled long enough to be falsely classified as dropped-out by our definitions.
    
    \item Deceased students. We exclude their data as well because death, formally meeting the definition of drop-out, does not meaningfully represent any scale of educational success.   
\end{enumerate}

\subsubsection{Brightspace (D2L)} 
Brightspace learning management system is a cloud-based software suite developed by the global software company, Desire2Learn (D2L). The pervasive popularity of Brightspace among the majority of American postsecondary institutions is owed to the facilities it provides for online and hybrid courses. However, some of these facilities, such as discussion forums, question banks, and grading tools, have attracted many instructors of traditional face-to-face courses as well.

The relevant information provided by Brightspace, which we directly collected through its API, includes data such as inner semester grades, quizzes' results, assignments, attendance, online access to materials, online discussions, and chats, which enable us to track students' activity index (indicating students' engagement in a course) as well as their performance during online courses. Our rationale for giving significance to student engagement has roots in the fact that, as we mentioned in Section 2, a strong positive correlation between engagement and performance in online courses has been fairly established. While some studies stress the short-term effect of engagement on performance, it is also suggested that engagement early in the semester is even more influential. This encouraged us to gather data from Brightspace on a weekly basis. 

Features represented by the data that we gather from Brightspace fall into two classes, namely inner-semester grades and results of online courses (7 features) and activity index (7 features). For the detailed distribution of these features, see Table \ref{tab:Compare} 

\subsubsection{National Student Clearinghouse (NSC)} 
We also gathered data from the student-tracker of The National Student Clearinghouse (NSC). NSC is a nonprofit organization that collects semester-based reports from nearly 3600 colleges and universities, covering 97\% of all students in public and private U.S. institutions. The data NSC provides regarding every student in each semester includes the name of the institutions students have enrolled in, the first and the last date of attendance, the type of enrollment (e.g., full-time, part-time, and withdrawn), and if the student has graduated in that semester. In case of graduation, the degree title and the graduation date will be given as well\cite{Voghoei_Decoding}. These reports enable us to trace students before and after enrolling in a specific school. Particularly, NSC's data helps us determine each student's status in terms of the labels mentioned above; especially, it facilitates distinguishing transfer students from ones who have dropped out. It is important to note that NSC data, as discussed later, further serves to differentiate two types of students within the category of transfer: the ones who graduate after transferring versus those who never graduate. 

\subsection{Centralization of Data}
All the data gathered from the resources above should be stored in a central database. However, the original data received from these resources had different formats of CSV, JSON, or Database. To remedy this inconsistency, we designed a pipeline to uniformly restructure the data in a central SQL-server database.

In order to extract data from Banner, we used PowerShell scripts that deliver this task on a daily basis and store the data in separate schemes within the central database so each scheme contains a distinct class of data (e.g., personal or financial). As for Brightspace, the same responsibility is to be fulfilled by a set of Python codes and Database triggers that extract data through Brightspaces's API and store them in a single scheme. The scattered data related to one student, separately gathered from these two resources, will be matched based on the combination of the first name, the family name, and the student's email address.     

With respect to the National Student Clearinghouse (NSC) a different procedure is required. For records related to each desired period of time and each specific set of concerning students, a separate request should be submitted to NSC. In response, NSC provides a CSV file presenting the demanded records. This part of the pipeline is handled by Python codes, PowerShell scripts, and Pl-SQL. These scripts, while processing the reports of NSC, are also responsible to react to two types of nuanced occasions: 1) the records of students who have spent a limited period of time at a different school between two enrollments in their original institution, 2) the records of students who were simultaneously enrolled in two institutions (co-enrollment). On both these occasions, we delete the data related to any institution other than the original one, because, given the definitions of the three labels we are concerned with, these data obviously do not contribute to labeling cases they represent. 

\section{Pre-Processing}
In our treatment of data, we started by excluding two types of samples from our training dataset:
\begin{enumerate}
    \item students continuing their education at the university of interest, due to not yet having fallen under any of the labels our model is intended to predict. 
    
   \item students who have left the university of interest within the last two years before 2019 and have not reappeared in the education system yet. The reason for the exclusion of this group is that they do not fit in any of the definitions given earlier for the labels.

\end{enumerate}    

In order to pre-process the data representing the samples that survived these exclusions, we have taken the following steps.

\subsection{Cleaning}
In the case of many students, the data suffers from inconsistency and imbalances. For example, a remarkable number of students have declared different races or ethnicities in different forms (e,g. FAFSA vs. application form), or at different times (e.g., two FAFSA forms submitted in two different years). On such occasions, we selected the most frequent value in student's records. When the frequencies are equal, the winning value would be selected randomly. However, if the nature of the data allows the possibility of its alteration over time (e.g., parents' highest academic degree), it will be included among time-series data, instead of fixed data.       

Other instances that should be the subject of cleaning are duplicated data and the records of co-enrollments, which one may find in abundance among NSC records. Co-enrollments happen when a student is simultaneously enrolled in the university of interest and another postsecondary institution. Although co-enrollment deserves to be studied as a potentially impactful feature, in the present stage of our study, we preferred to ignore such records. Nevertheless, including this feature in our model in the future stages of the study, will be imperative.

\subsection{Missing Values}
Due to the abundance of missing values in our dataset, dealing with them has become one of the most challenging problems in our project. In general, we have adopted three strategies to address this challenge:

\begin{enumerate}
    \item In the cases in which missing a value that would represent a feature bears interpretable or accountable information, we group students based on lacking such specific features to be studied by a separate model. For example, for students who have not applied for FAFSA, no value represents features such as individual or family financial status. However, we find it reasonable to assume that having not applied for FAFSA implies some extent of financial sufficiency, which allows us to group these students.   
    
    \item There are features that, when no value is available for which, their natures let us decide either to represent them with zero, to impute them with the means or the median of the distribution of existing values, or to use k-Nearest Neighbors (KNN) to predict the missing values based on correlated features. While attempted credit hours and GPA are examples apt for the first options, age and family income lead us to the second and third options.
    
    \item Regarding features for which more than 80\% of samples do not return any value, we decided to remove the feature from our database. The reason is that we did not find imputing such features based on other data a trustworthy approach. Although most of the high school scores, like ACT English (A01) and ACT Math (A02), and some features reflecting family backgrounds belong to this category, we decide to remove them from the dataset as long as we do not have access to a better resource to collect data representing these features for a big enough number of students.

\end{enumerate}
\subsection{Outliers}
Our approach towards outliers, in the first place, depends on how possibly they represent the matter of fact. If they are not possible (e.g., 5 or 205 for the age of a student), then, depending on the nature of the feature, they will be replaced with the mean or the median or the z-score of the values representing the feature in the dataset. 

In the case of possible outliers, we distinguish between the features whose values should be interpreted continuously and the ones whose informative values are interpreted discretely. We apply the interquartile rule to identify and replace the former with the max or min of the rest of the values. For the latter, we practice data binning to make these values are not left unrepresented.

\subsection{feature engineering}
Although the raw data extracted from the resources mentioned earlier directly represent many significant features, many relevant classes of information need to be engineered. For example, while the GPA for a specific semester can be directly extracted from the data resources, the overall GPA needs calculating. Age and driving distance between the residence and the school is also among the engineered features. The former is calculated based on the date of birth, and the latter, which requires a more complicated procedure, is engineered based on the Geocode addresses of students provided by the Banner.  

Using data like marital status, the number of children, or salary, which are expected to change over time, in addition to including them in time-series data, we engineer features like the number of times they have altered and the max and min of their values, if applicable.         

As examples of engineered features related to online courses, we can mention the following: the total time each student spends on each course (or totally on all courses in a semester), weighed grades, and most of the activity indices like the number and the total time of sessions in which students view course materials and the number of their posts, messages, and quiz attempts.

\subsection{Converting Categories to Numeric Values}
As will be discussed later in this paper, the models we used for this study are Deep Learning. This required us to convert all feature values expressed by discrete categories to numeral values. For this purpose, to avoid imposing the quantitative nature of numeral values on categorical classifications, we used the One-Hot in the case of many features like marital status, race, and gender. 

\subsection{Centralization and normalization}
To make sure that all the features are represented in the same order of magnitude, which is suitable for a comprehensive comparative evaluation, we used RobustScaler. It centers and scales statistics based on percentiles and, consequently, protects the model against the influence of a few outliers. Also, we used a normalizer to re-scale the vector of each feature, which maintains a harmonized dataset. Improving the conversion speed is expected to help the model find the optimal solution.

\subsection{Bias}
Recently the machine learning community has witnessed examples of bias that have hit the headlines. Most supervised learning algorithms, utilized toward decision making in everyday applications, may at first be presumed devoid of intrinsic biases, but in fact, inherit and reflect any bias or discrimination present in the training data \cite{ntoutsi2020bias}. Overall, there are some data quality issues that contribute to bias in ML applications. These issues may include insufficient data on certain categories (i.e., too few examples overall for certain minority groups), or incomplete representation of some groups due to a surfeit of missing values. Toward a fairer and less discriminative decision making, we use some pre-processing techniques to mitigate and reduce bias. We identify the features which are more likely to be bias-contributing and select relatively similar sample sizes from each category. Proxy features which are those features with a high correlations with the target and the biased outcome, such as the zip code are avoided in our treatment of the data or engineered features. Alternatively, in some cases we utilized the sampling or weighing techniques which would save the core of the data but would create less bias. However, tackling bias in data is a challenging task which could always use some improvements. 

\subsection{Balance}
In reality, students are not equally distributed across the four categories we are concerned with. While in the first semesters, more students fall in the category of transfer, the longer students stay in school, the more the balance leans in favor of graduation (see \ref{fig:LabelsDist}). This imbalance allows the majority to dominate the results. To feed the balanced data into the model, we applied two simple methods, namely, Standard Equal Random Selection and MICE. At the end, MICE proved to provide better results.     

\begin{figure}[hbt!]
\centering
    \includegraphics[width=0.5\textwidth]{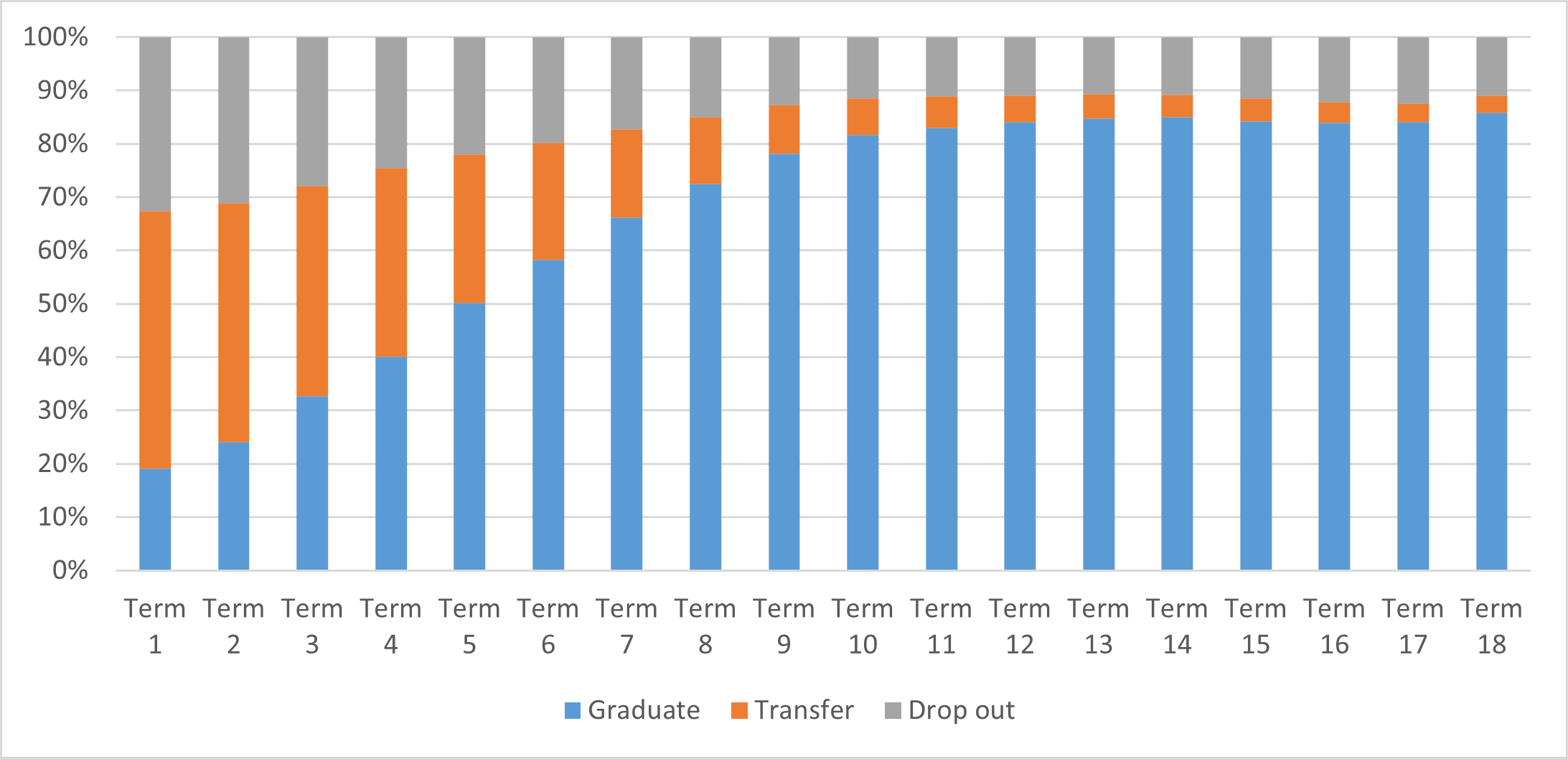}
    \caption{Labels Balance}
    \label{fig:LabelsDist}
\end{figure}

\section{Method}
As mentioned before, two sorts of features are included in this study: the features whose values are not expected to change during the period the model is concerned with (fixed data), and the ones whose values may change over this period (time-series data). While race, and to some extent, gender are examples for the first category, the number of attempted credit hours and the activity indices that are measured during online courses represent the second category. This fundamental divide in the dataset projects itself on the typology of the models we have incorporated into our method. To handle the fixed data, CNN and DL have been alternatively used and in the end, DL proved to be more capable of identifying graduate students and dropouts, while CNN turned out to be more successful at labeling transfer students. Finally, handling time-series data has been entrusted to LSTM. At the end, the outputs of LSTM, DL, or CNN have been concatenated to be followed by layers of FC-DL to learn the outcome smoothly.

\begin{figure*}[hbt!]
\centering
    \includegraphics[width=1.8\columnwidth]{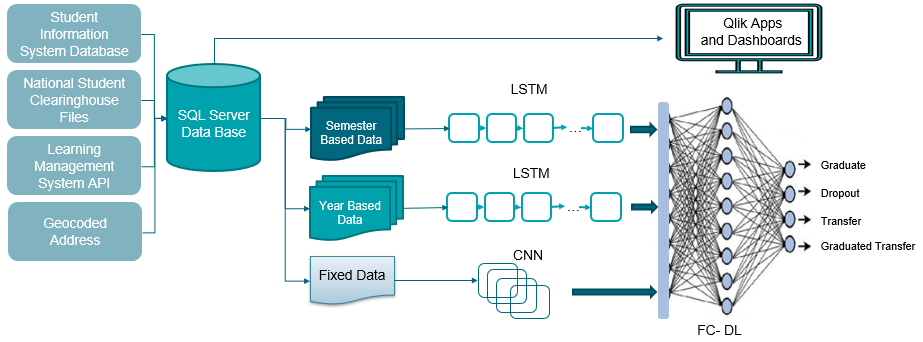}
    \caption{The model Structure}
    \label{fig:ModelStructure}
\end{figure*}

Our study undertakes to label students in the following stages of curricular progress:

Stage 1: The end of the add-drop period of the first semester;

Stage 2: The end of the first year;

Stage 3: The end of the second year;

Stage 4: The end of the third year.

Since, for Stage 1, the whole data is fixed, there is no need for the LSTM. Therefore, CNN and DL are sufficient to constitute the models. At the end of the first year (Stage 2), an LSTM will be required to handle the time-series data collected during the last two semesters, while the fixed data is still dealt with by CNN and DL separately. For the second and the third years (Stages 3 and 4), in addition to the LSTM handling semester-based data (four semesters for the second year and six semesters for the third year), a second LSTM also is needed to work with the time-series data that have been collected yearly (e.g., FAFSA data is collected once every year). Again, the output of all these LSTMs will be concatenated with that of CNNs and DLs, alternatively. Figure (\ref{fig:ModelStructure}) presents an overall sketch of the models.

In every dataset used in this study, for the purpose of testing, we randomly reserved 2\% of the sample set as unseen data; 2\% of the sample set also was reserved for development; and the rest was the subject of training.

\section{Discussion and Results}

We originally considered four labels to be predicted by our models, namely, Graduate(G), Dropout(D), Transfer to-be-graduate (hereafter addressed as Transfer-G or TG), and Transfer(T) i.e., the ones who transfer but their records do not show any graduating. However, educators, based on motives such as the interest of the institutions they serve or the limitations of resources, may demand to adjust our method for different scenarios in which two of these categories collapse together or the direction of transferring (transfer-up vs transfer-down) is specified. For example, the resemblance between the learning behaviors of dropout students and the ones that will drop out after transferring may persuade advisors to group them together. In another scenario, having Dropout, Transfer, and Transfer-down fallen together vs. the cluster of Graduate and Transfer-up, may be informed by a redefined notion of at-risk students. These scenarios will be worth considering in our study, especially if they improve the precision of our models. Therefore, in addition to the four-label scenario, we decided to adjust our models for some alternative scenarios that are as follows:

\begin{figure}[hbt!]
\centering
    \includegraphics[width=1\columnwidth]{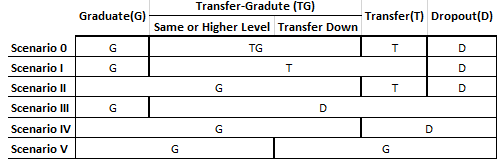}
    \caption{Distribution of Labels in Five Scenarios}
    \label{fig:scenarios}
\end{figure}

Scenario 0: In this scenario, the labels are Graduate, Dropout, Transfer-G, and Transfer (four labels).

Scenario I: In this scenario, Transfer-G collapses with Transfer, which stands in contrast with Graduate and Dropout (three labels).    

Scenario II: In this scenario, Transfer-G collapses with Graduate, which stands in contrast with Dropout and Transfer (three labels). 

Scenario III: In this scenario, Transfer-G and Transfer collapse with Dropout and stand in contrast with Graduate (two labels). This scenario represents the traditional mode of classification with which previous studies have been concerned. 

Scenario IV: In this scenario, Transfer-G collapses with Graduate, and Transfer collapses with Dropout, so that, finally, only Graduate and Dropout stand against each other (two labels).

Scenario V: We also experimented with a variant of Scenario IV, where we mean Transfer-G to represent only the transfer students who eventually graduated in a program of the same level (or higher) as they left. It will be the only example in our study in which the direction of transfer matters.

Figure \ref{fig:scenarios} compares the areas each category covers in the scenarios mentioned above.

In the following sub-section, we will present the best achievements of our experience with the models with respect to the aforementioned scenarios. In order to reveal more practical aspects of our methods, we will also specify the scenario in which the model always demonstrates acceptable performance, the stages after which the model identifies all the categories with a trustworthy degree of accuracy, and the stage by which the data has become suitable enough for making useful predictions in a specific scenario. 

\subsection{Performance}

\begin{figure*}[hbt!]
\centering
    \includegraphics[width=1.5\columnwidth]{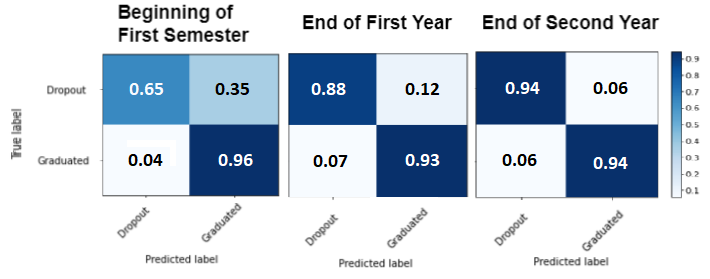}
    \caption{Beginning the first semester by overall accuracy of 80.25, the end of first semester by 90.13 and the end of second semester by 94.46}
    \label{fig:2Labels}
\end{figure*}

\begin{figure}[hbt!]
\centering
    \includegraphics[width=1\columnwidth]{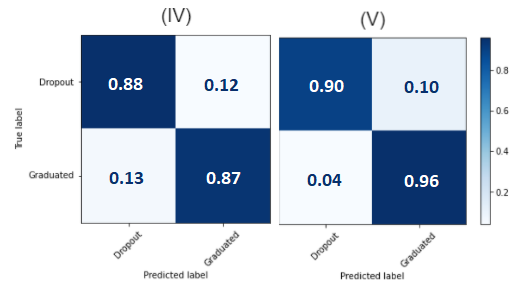}
    \caption{Scenario IV show more confusion when we collapse the transferred Graduate student into Graduate compare to when we classified them as dropout, and big improve in scenario V when we consider graduate only those graduated in the same level or above  }
    \label{fig:2LabelsScenario}
\end{figure}

Before everything, it is to mention that, in the traditional scenario (Scen. III), which has been the scenario in almost all previous studies, our models display a reasonably high performance. By the time a freshman has finalized the courses they would take in the first semester, our model identifies to-be-graduate students and the ones the university is going to lose (dropouts and transfers) with 80.25\% accuracy. This accuracy will increase to 90.13\% and 94.46\% at the end of the first and second years, respectively (see Figure \ref{fig:2Labels}). If in a binary scenario, by Graduate we mean whoever finally has graduated somewhere in the education system (the union of Graduate and Transfer-G in contrast with the union of Transfer and Dropout, which defines Scen. IV), the accuracy will drop (e.g., 87.54\% at the end of the first year; see Figure \ref{fig:2LabelsScenario} - left). However, if in such a situation we redefine Transfer-G as the transfer students who graduate in a program not lower than the one they have transferred from (Scen. V), the accuracy will improve (e.g., 93.06\% at the end of the first year; see Figure \ref{fig:2LabelsScenario} - right). We account for this improvement by hypothesizing a significant resemblance between the features of graduates and transfer-up students, on the one hand, and the same between those of dropouts and transfer-down students, on the other hand. We believe that these similarities confuse the models when, in Scenario IV, all transfer students, regardless of their fates, are classified together with dropout students. 

\begin{figure}[hbt!]
\centering
    \includegraphics[width=1\columnwidth]{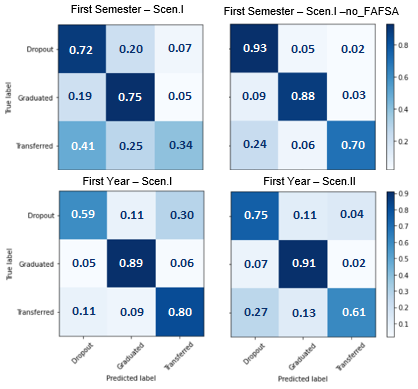}
    \caption{Recalls at the Beginning of the First Semester (Scen. I and II)}
    \label{fig:3Labels}
\end{figure}

Although in the beginning of the first semester, we were fairly able to recognize the students that the university of interest would lose one way or another, this stage is not the perfect time to make a better resolution within this group of students. However, at this stage, the model returns 83.82\% accuracy for Scenario I (with 3 labels; see Figure \ref{fig:3Labels} - top right), provided it runs only with students for whom there is no FAFSA records. Especially, among this group of students, it has succeeded in identifying 93\% of Dropouts with F1-score = 82\%. This is a remarkable performance in comparison with the case in which all students with and without FAFSA have been learned together (accuracy = 60.55\%; see Figure \ref{fig:3Labels} - top left). Interestingly, we also noticed a slight improvement when we ran the model only for no-FAFSA students in this stage, with Scenario 0. In “Future Work”, we will come back to this observation and what it may imply. 

The data accumulated by the end of the first year contributes to the resolution of our predictions. In this stage, in addition to all abilities that binary scenarios offer to us, we will be able to identify 80\% of all students likely to transfer (regardless of their fates), with F1-score = 74\% (Scen. I; see Figure \ref{fig:3Labels} - bottom left). In this stage, the model also can predict 75\% of the cases in which students drop out before transferring, with F1-score = 72\%, as well as 91\% of students who finally graduate somewhere in the education system, with F1-score = 85\% (Scen. II; see Figure \ref{fig:3Labels} - bottom right)

\begin{figure}[hbt!]
\centering
    \includegraphics[width=0.8\columnwidth]{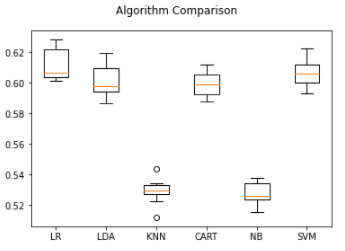}
    \caption{Performance of Six Models at the End of the Second Year (Scen. I)}
    \label{fig:ComparedModels_2Year}
\end{figure}

\begin{figure}[hbt!]
\centering
    \includegraphics[width=1\columnwidth]{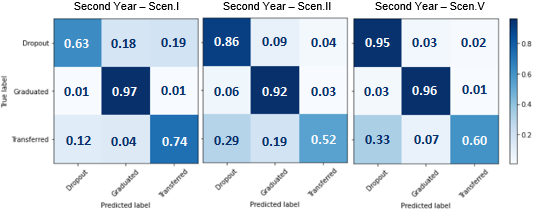}
    \caption{Recalls at the End of the Second Year (Scen. I, II, V)}
    \label{fig:2years}
\end{figure}

At the end of the second year, the dataset serves the model to classify students in Scenario I with 78.28\% accuracy (Figure \ref{fig:2years} - left). To demonstrate the advantages of our model over the ones usually used in similar studies, we have compared the performance of six models, namely, Logistic Regression (LR), Linear Discriminant Analysis (LDA), K-Neighbors Classifier (KNN), Decision Tree Classifier (CART), Gaussian Naive Bayes (NB), and Support Vector Machine (SVM) on the dataset for the same scenario at the same stage in Figure \ref{fig:ComparedModels_2Year}.

At the end of the second year, the model returned an accuracy of 76.65\% for Scenario II, identifying 92\% of to-be-graduate and 86\% of dropouts, with F1-scores 83\% and 78\%, respectively (Figure \ref{fig:2years} - center). This will increase to 83.06\%, provided we have graduate students and the ones that transfer and graduate in a program not lower than the initial program collapsed together (Scen. V). In this scenario, the model identifies 96\% of to-be-graduate students and 95\% of dropouts, with F1-scores 93\% and 82\%, respectively (\ref{fig:2years}  right).

\begin{figure}[hbt!]
\centering
    \includegraphics[width=1\columnwidth]{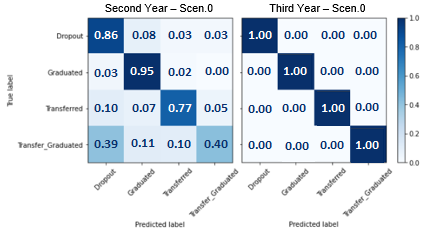}
    \caption{Recalls at the End of the Second Year and Third Year (Scen. 0)}
    \label{fig:4Labels}
\end{figure}

\begin{figure}[hbt!]
\centering
    \includegraphics[width=0.8\columnwidth]{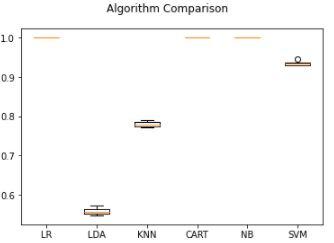}
    \caption{Performance of Six Models at the End of the Third Year (Scen. 0)}
    \label{fig:ComparedModels_3Years}
\end{figure}

A resolution as good as that of Scenario 0 is achievable only at the end of the third year. In this stage the model identifies students belonging to all four categories of Scenario 0 with 99.34\% accuracy (Figure \ref{fig:4Labels} - right). In order to make sure this result is not unrealistic, we tried the six models mentioned in the previous paragraph for this case. Their results, which support the performance of our model, are demonstrated in Figure \ref{fig:ComparedModels_3Years}

\subsection{Impactful Features}

Our study undertakes to identify the most important features in the procedure of prediction, in different terms. 

\begin{figure}[hbt!]
\centering
    \includegraphics[width=1\columnwidth]{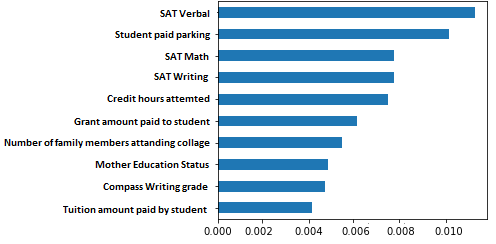}
    \caption{Impactful Features to the Classification of Students in Scenario I (Beginning of the First Semester}
    \label{fig:Top10_sem1}
\end{figure}

As early as the beginning of the first semester, we can make the list of the most impactful features on the overall result of our models in a given scenario. For example, SAT scores, whether the student pays for parking, the number of attempted credit hours, the amount of grant the student has received, family's contribution to tuition fee payment, and mother's education are the features that make the heaviest contributions to the classification of students in Scenario I, right in the beginning of the first semester (see Figure \ref{fig:Top10_sem1}).

\begin{figure}[hbt!]
\centering
    \includegraphics[width=1\columnwidth]{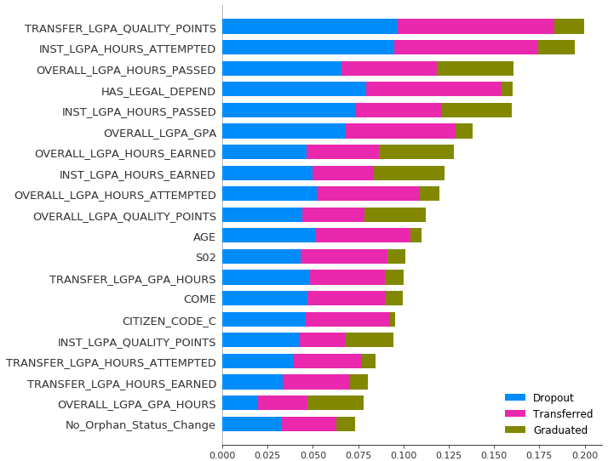}
    \caption{The Ratios of the Contributions of Top Selected features to Each Label}
    \label{fig:lastTopFeatures}
\end{figure}

The model also can specify the ratios of the contributions of these features to each label. Figure \ref{fig:lastTopFeatures} gives an example of such specifications with respect to Scenario I at the end of the second year. It is easy to notice that the list of the most impactful features in this example is very different from the one made for the beginning of the first semester. 

\begin{figure*}[hbt!]
\centering
    \includegraphics[width=1.8\columnwidth]{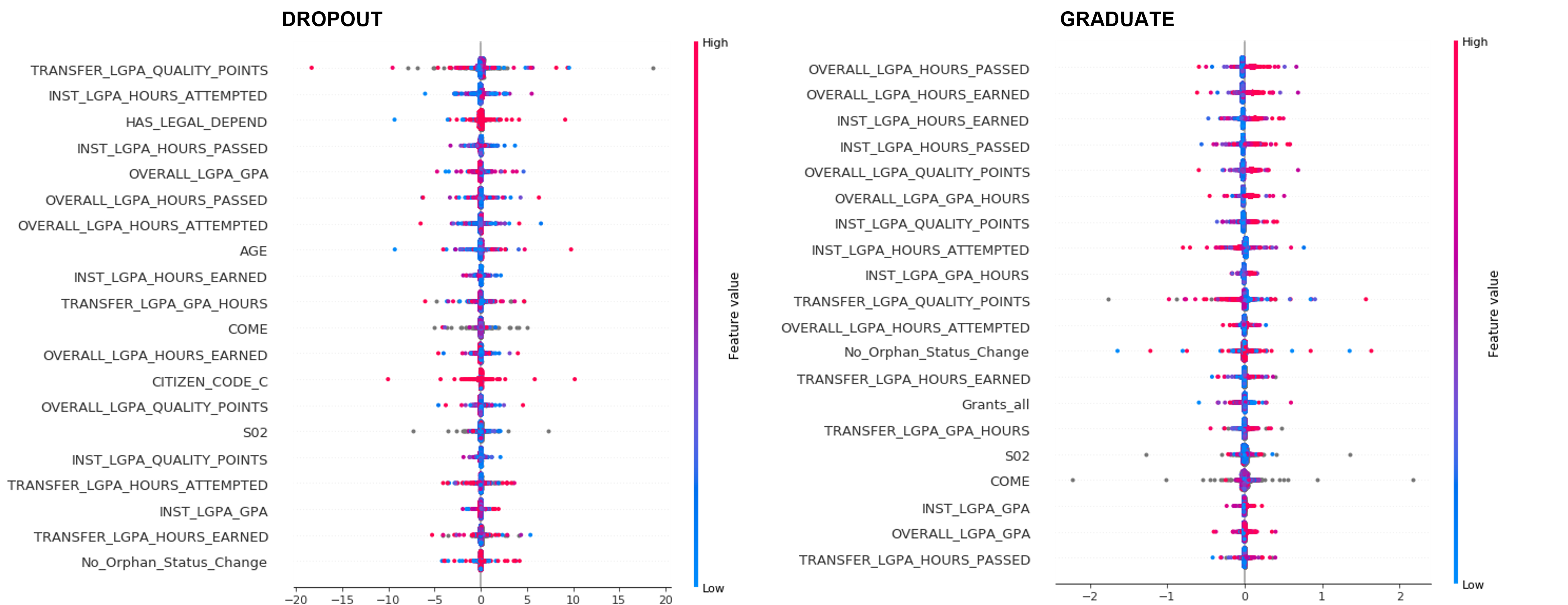}
    \caption{Most Important Features for Each Lable}
    \label{fig:GradVSDrop}
\end{figure*}

Furthermore, the model enables us to recognize the features that contribute to the prediction of each label independently. It gives an opportunity to educators to find the most important factors that may cause students to end up in undesired situations, among the features suggested by our model. In Figure \ref{fig:GradVSDrop}, the most important features contributing to graduation and dropout have been sorted independently (for Scenario I, at the end of the third year).

\begin{figure*}[hbt!]
\centering
    \includegraphics[width=1.8\columnwidth]{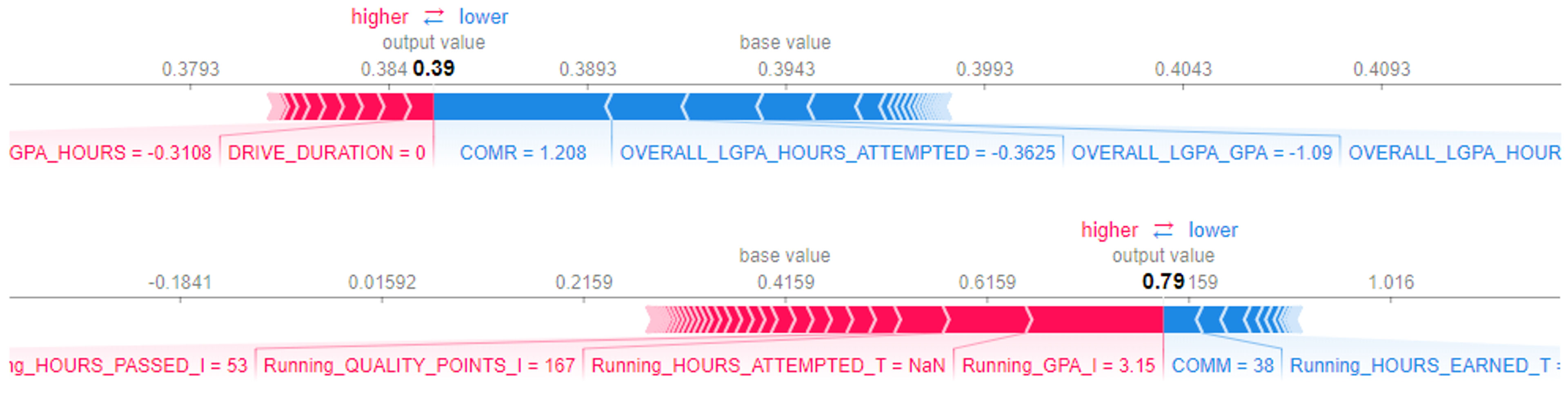}
    \caption{Personalized Report: The Impact of each Feature for each Student}
    \label{fig:FeatureIndividual}
\end{figure*}

As a remarkable achievement, this model is the first of its kind that is capable of personalizing the list of the most contributing features in labeling an individual student, sorted by the intensity of impact. Figure \ref{fig:FeatureIndividual} represents such sorting for two manufactured samples. 

We are also able to deliver personalized reports with respect to each student, for a given stage and a given scenario. In these reports, we will present the likelihood that the student ends up falling in each category as well as the most impactful features in this personalized prediction, sorted by their weights. Table \ref{fig:FeatureIndividual} illustrates the format of such reports for some manufactured samples. This format has been designed based on the expectations of the university of interest and can be modified to satisfy the demands of other schools.  

\begin{figure}[hbt!]
\centering
    \includegraphics[width=1\columnwidth]{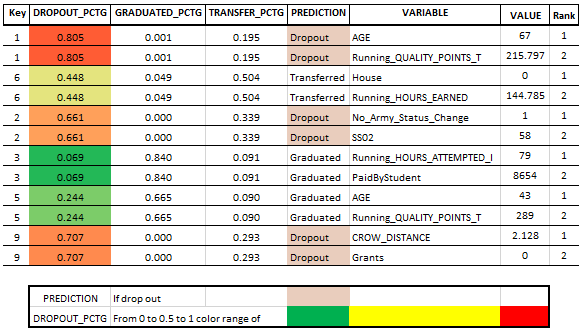}
    \caption{The deliverable file structure in the student level. This figure only show 2 top features per student. Values are fake! }
    \label{fig:output}
\end{figure}

\section{Future Work}
One of the main concerns of any further attempt to improve the present study should be the task of feature enhancement. There are categories of information to which we could not find any access. Many of these categories have been suggested by the researchers of education science as the factors that significantly influence students' educational performance. Changing majors, having student jobs like teaching assistantships, being on probation, belonging to the first generation in the family who attends college, speaking English as a first language, the intensity of engagement with libraries, and the reason for withdrawal if there has been any, are among the factors supported by the literature of which we did not have any records for different reasons. For example, although the library of the university of interest keeps records of books that are checked in and out, these records are not connected with Banner, so it would not be possible for us to include them in our study. To use features like the reason for withdrawal, we will need to create a text mining model to extract the data and convert it to a format consumable for the model we discussed in this paper. Furthermore, features like the ones related to students’ first language should be considered in data gathering methods in the future.

Given that the most unfavorable ending for a student and the entire education system is a dropout, we assume that the precise identification of at-risk students would be the first expectation of advisors. In this respect, the more students predicted to graduate end up dropping out, the more the trust of advisors in our model is betrayed. However, until the end of the second year, our model, in several runs, kept labeling 12\% of students who are destined to drop out as to-be-graduated. It suggests that a portion of students share enough features with students to-be-graduate, but, for reasons undetectable for our model, leave the school for good. We found this suggestion acceptable for the expert educators we consulted. To fulfill the mission undertaken by our study, we need to enlarge the set of our features such that it finally covers the distinctive characteristics of this portion.    

Our observations in the case of FAFSA also offer suggestions worth considering in the future developments of this study. As we mentioned earlier, FAFSA is a unique source of values representing many features. Our models have identified some of these features among the most impactful factors. However, a good number of students (nearly one-third) have never filled a FAFSA form. Surprisingly, we observe that when our models are trained separately for students who have FAFSA data and the ones who have not, the accuracy associated with the latter group surpasses the one associated with the former. We account for this phenomenon by blaming the plenitude of missing values in FAFSA records that supposedly confuses the models trained with FAFSA. It implies that enforcing the classes of information similar to those of FAFSA in future methods of data gathering should be expected to contribute to the performance of models like ours.

Finally, our model is not able yet to identify the destination programs of transferring students. If, for example, a school comes to know that most of the students who transfer from its Bachelor's programs join Certificate programs somewhere else, it may be able to keep its students by increasing the variety and capacity of its Certificate programs. The future development of our study will make answering this question one of its missions. 

\section{Conclusion}

To design a model capable of classifying students based on the likelihood of their ultimate educational persistence and achievement, we conducted a study on the records of 63,917 students enrolled in a university in the state of Georgia, which we address as the university of interest. From these records, which reflect students' information from 2006 to 2019, we extracted data that finally maintained 121 features to feed into our model. These features are divided into two categories, which will be dealt with separately: the fixed data and time-series data. We insisted on collecting the data from the learning management system and the resource planning software used by the university of interest, and the nationwide tracking system of the National Student Clearinghouse (NSC). To classify students by their likelihood of falling into categories corresponding to their fates in the educational system, in general, and at the university of interest, in particular, we made a model in which the output of one or more LSTMs trained on the time-series data and that of a Deep Neural Network trained on the fixed data are concatenated. An alternative model was made in the same fashion, except in which the Deep Neural Network was replaced with a CNN. While, in the end, the former turned out to be more precise at identifying dropout students, the latter achieved higher overall accuracy and had a better performance in the case of graduates. It is worth noting that our study is among very few that have addressed the understudied challenge of labeling students likely to transfer.While researchers have typically failed to differentiate between different forms of attrition (e.g.,\cite{stenton2021fine}), our model distinguishes transfer from drop out and, more importantly, differentiates transferring up to a better school from transferring down to a lower-ranked college.

Addressing different possible purposes for which a school may need to define and identify at-risk students, we provisioned a number of scenarios for classification based on whether or not transfer students in general, transfer students who finally do not graduate, or the ones who transfer to a lower-level program should collapse together with dropouts. For each of these scenarios, we ran several instances of our models trained with data collectible by the end of four stages of data gathering, namely, the add-drop period of the first semester of attendance, the end of the first year, the end of the second year, and the end of the third year.

In the traditional binary scenario, which is the only scenario most of the previous studies were concerned with, our model returns a remarkable performance (80.25\% accuracy as soon as students enter the college, which increases to 94.46\% by the end of the second year).  Our method also, provided with the data collected in the first couple of weeks of the first semester, succeeded in identifying 93\% of non-FAFSA students who were likely to drop out from the university of interest (vs. the ones who would transfer or graduate from that university) with F1-score = 82\% (overall accuracy of the model = 83.82\%). By the end of the first year, the collected data enables us to classify the students who are likely to graduate somewhere in the education system with an overall accuracy of 87\%. This accuracy will increase to 93.06\% when students who transfer and graduate in a program lower than the original one are not supposed to represent educational success. At the end of the second year, the model can identify 92\% of to-be-graduate students with F1-score = 83\%, vs. 86\% of dropouts, with F1-score = 78\%. In this stage, when the level of transfer is involved in the definition of graduation, the model labels 96\% of to-be-graduate students (F1-score = 93\%) vs. 95\% of the rest (F1-score = 82\%) with an overall accuracy of 83.06\%. However, by the end of the third year, as expected, the model succeeds in classifying students, in a scenario with the highest resolution, with an overall accuracy of 99.34\%.

Our models also single out the features that are the most impactful in our predictions (for each stage/scenario) and sort them by the intensity of their impacts. However, what we consider as one of the most important achievements of this study is that our study is the first of its kind that identifies such important features and specifies the proportion of this contribution for each student. We hope this achievement provides advisors functioning on individual scales with a game-changing insight into the causal net of factors that should be altered to avoid undesired endings in the educational journey of their students.

\section*{Acknowledgment}
The authors would like to thank David Tanner, and John A. Miller for their advice and support. 

\section*{References and Footnotes}

\subsection{References}
\bibliographystyle{unsrt}
\bibliography{main}

\subsection{Footnotes}
\section{Features Descriptions}
\onecolumn
\begingroup
\small
\begin{longtable}[c]{| c | c |}
 \caption{Glossary of Abbreviations Standing for Features \label{long}}\\

 \hline
 \hline 
 Abbreviation &  Extension\\
 \hline
  \endfirsthead
  
  \hline
 \multicolumn{2}{|c|}{Continuation of Table \ref{long}}\\
 \hline
 Abbreviation &  Extension\\
 \hline
 \endhead

 \hline
 \endfoot
 \hline 
A01 & ACT ENGLISH\\
A02 & ACT MATH\\
A03 & ACT READING\\
A04 & ACT SCIENCE REASONING\\
A05 & ACT COMPOSITE\\
ACCM & ACCUPLACER CLASSIC ELEM. ALG.\\
ACCNGM & NEXT GEN ACCU QR/ALG/STATS\\
ACCNGR & NEXT GEN ACCU READING\\
ACCR & ACCUPLACERCLASSIC READINGCOMP\\
ACCW & ACCUPLACER WRITEPLACER\\
Aids & Total financial aid\\
AT-RISK-HOMELESS & If student is at the risk of being homeless\\
Avg-Points & Average points from online classes\\
Avg-Points-perClass & Average points in semester\\
Avg-Weighted & Average weighted points from online classes\\
Avg-Weighted-perClass & Average weighted points from online class\\
COME & Indicate the Compass Writing grade\\
COMG & Indicate the Compass Geometry grade\\
COMM & Indicate the Compass  Algebra grade\\
COMR & Indicate the Compass Reading grade\\
CROW-DISTANCE & CROW distance\\
DEPEND-AGE-0-5 & Number of student's dependence (age less than 5)\\
DEPEND-AGE-13-PLUS & Number of student's dependence (age greater than 13 )\\
DEPEND-AGE-6-12 & Number of student's dependence (age between 6 and 12)\\
DRIVE-DISTANCE & Driving distance from living place to school\\
DRIVE-DURATION & Driving  time from living place to school\\
FAM-MEMB & Number of people in the student's household\\
Father-EDU-Status-Changed & If the father's education status changed?\\
FATHER-HIGHEST-GRADE & student's father's education status\\
FOOD-STAMPS & If student used the food stamp\\
GPA & The student GPA for the current semester\\
Grants & Total grant amounts\\
HAS-LEGAL-DEPEND & If has legal dependents other than a spouse\\
HOURS-ATTEMPTED & Hours student attempted\\
HOURS-EARNED & Hours student earned\\
House & Indicate if the student owns a house\\
HS-GPA & High School GPA\\
I-GPA & Institute GPA\\
I-HOURS-ATTEMPTED & Institute hours student attempted\\
I-HOURS-EARNED & Institute hours student earned\\
I-QUALITY-POINTS & Institute quality point\\
Loans & Total amounts of loan student got\\
LSE & Learning Support English Grade\\
LSM1 & Learning Support MATH 1 Grade\\
LSM2 & Learning Support MATH 2 Geade\\
Meal & Indicate if the student used the meal plan\\
Mother-EDU-Status-Changed & If the mother's education status changed?\\
MOTHER-HIGHEST-GRADE & Student's mother's education status\\
No-ForumId & Number of forum student participated in \\
NO-IN-COLL & total number of family members attend college\\
indicate  & The number of post student sent per semester\\
No-Session & Total session time\\
No-Submission & Total number of assignment submision\\
No-TopicId & Total number of topic student participated in\\
Paid By Student & Total amount which student paid from his/her tuition \\
PAR-FAM-MEMB & The number of family members in the parent's household.\\
PAR-FOOD-STAMPS & If anyone in the parents' household received SNAP\\
PAR-MRTL-STATUS & The marital status of the student's parents.\\
PAR-NO-IN-COLL & Number of parent's household in college\\
PAR-RECD-SSI & Parents received Supplemental Security Income.\\
PAR-RECD-TANF & Parents received Temporary Assistance\\
PAR-RECD-WIC & Parents received assistance for Women Infants and Children\\
PAR-SCHOOL-LUNCH & Indicates  if parents received Price School Lunch benefits.\\
Parking & Indicate if the student used parking?\\
PELL & Total amounts of PELL student got\\
QUALITY-POINTS & Quality point per semester\\
RECD-SSI & RECD-SSI\\
RECD-TANF & RECD-TANF\\
RECD-WIC & RECD-WIC\\
S01 & SAT CRITICAL READING\\
S02 & SAT MATHEMATICS\\
S05 & SAT TSWE SCORE\\
S07 & SAT WRITING\\
SPS-INC-FR-WRK & The amount of income earned from work \\
SS01 & SELF REPORTED SAT VERBAL\\
SS02 & SELF REPORTED SAT MATH\\
SS07 & SELF REPORTED SAT WRITING\\
Sum-AttemptNumber & Total number of quiz attempts\\
Sum-NumViews & Total number of course material views\\
Sum-SessionSecond & Total session time\\
T-GPA & Transferred GPA\\
T-HOURS-ATTEMPTED & Transferred hours student attempted\\
T-HOURS-EARNED & Transferred hours student earned\\
T-QUALITY-POINTS & Transferred quality points student earned\\
TOT-FAM-CTRB & The total family contribution\\
TOT-SAR-EFC & what the SAR EFC was for a person\\
Tuition-Fee & The total amount of Tuition fee \\
US-VET & Indicates if the student is in US Armed Forces.\\
Visit-AvglTime-Class & Average visit time for class material \\
Visit-TotalTime-user & Total visit time for class material \\
WARD-OF-COURT & Reported a ward of the court or is an orphan.\\

 \end{longtable}
 \endgroup
 \twocolumn

\EOD

\end{document}